\documentclass[letterpaper]{JHEP3}
\title{Spacetime dynamics and baryogenesis in braneworld}
\author{Tetsuya Shiromizu \\ 
Department of Physics, Tokyo Institute of Technology, 
Tokyo 152-8551, Japan, \\
Advanced Research Institute for Science and Engineering, 
Waseda University, Tokyo 169-8555, Japan
\\ \email{shiromizu@th.phys.titech.ac.jp}}
\author{Kazuya Koyama \\  Department of Physics, The University of Tokyo, Tokyo 
113-0033, Japan
\\ \email{kazuya@utap.phys.s.u-tokyo.ac.jp}}

\abstract{
We point out that the effective theory for the Randall-Sundrum braneworld models with 
bulk fields contains the baryon number violation process depending on the 
spacetime dynamics. Combining to the curvature-current interaction,  
the net baryon number observed today may be explained. 
The resultant baryon to entropy ratio is determined by the ratio of the 
Planck scales in four dimensional and five dimensional spacetime 
except for the parameter for CP violation.
}
\begin{document}
\section{Introduction and summary}

The origin of the baryon asymmetry is an important problem in cosmology \cite{Review}. 
Recently, stimulated by the development of string theory, the brane world 
idea is actively investigated. In the brane world, a new mechanism to realize 
baryon number violation process arises \cite{DG}.  
In this paper we shall point out that there exists the classical baryon number (B) 
violation process in the Randall-Sundrum (RS) brane world model with bulk fields.  
Even if they have no potentials, the baryon number conservation is violated 
by the spacetime dynamics via the four dimensional curvature and its derivative terms. 
In order to explain the observed Baryon asymmetry, we consider the CP/CPT violation 
via the curvature-current 
interaction ($\partial_\mu R) J^\mu$. This interaction is also naturally 
induced on the brane if there is an explicit CP violating interaction. 
In order to make this mechanism work, $\dot{R} \neq 0$ is necessary in radiation
dominated (RD) universe. In four-dimensional general relativity $\dot{R}=0$ in 
RD universe. However, interestingly, $\dot{R} \neq 0$ is realized in the 
RS model due to the effect of the higher dimensional gravity. 
Then we show that the observed baryon to entropy ratio can be explained.  
Our finding indicates that the higher-dimensional gravity will bring a new 
way to generate baryon asymmetry in the universe via the dynamics of the spacetime.

\section{Baryon number violation from spacetime dynamics}
We begin with the action motivated by
the Randall-Sundrum(RS) model \cite{RS,BWreview}. 
%
\begin{eqnarray}
S_{\rm tot}=S_{\rm bulk}+S_{\rm brane},
\end{eqnarray}
%
where $S_{\rm bulk}$ is the bulk action 
%
\begin{eqnarray}
S_{\rm bulk}=
\int d^5x {\sqrt {-G}}\Biggl[
\frac{M_5^3}{2}\:{}^{(5)}R(G)-\Lambda-|\nabla_M \Phi|^2  \Biggr],
\end{eqnarray}
%
and $S_{\rm brane}$ is the brane action 
%
\begin{eqnarray}
S_{\rm brane}=\int d^4 x {\sqrt {-g}}\Biggl[-\sigma + {\cal L}_{\rm matter} \Biggr]. 
\end{eqnarray}
%
$G_{MN}$ is the five dimensional bulk metric and $g_{\mu\nu}$ is the brane induced metric. 
$\Lambda$ is the bulk negative cosmological constant and $\sigma$ is the brane tension. 
We assume RS  tuning and take $\Lambda = -\frac{\sigma^2}{6M_5^3}$.
$\Phi$ is the bulk complex scalar field. In adS spacetimes, the scalar fields is localised 
on the brane as graviton. 

We will derive the effective theory on the brane. 
Our argument relies on the braneworld holography. It is well known that 
the holographic picture is naturally held in RS models. 
The five-dimensional theory can be described by the four-dimensional theory 
coupled to the CFT. This is because the setup 
in RS models is quite resemble to that of adS/CFT correspondence \cite{GKR,SKOT}. 
For example, the zero mode($\varphi$) of the bulk complex scalar field($\Phi$) will 
be localised on the brane \cite{BG}. In our model, $\varphi$ might be regarded as 
the squarks or sleptons that carry the Baryon/Lepton number. Applying the braneworld 
adS/CFT correspondence \cite{GKR,SKOT} to the present model, we will be able to have the 
following effective action on the brane (See 
appendix A for the brief sketch of the derivation.) 
%
\begin{eqnarray}
S_{\rm eff} & \simeq &  \int d^4x {\sqrt {-g}} \Biggl[ \frac{M_4^2}{2}R+{\cal L}_{\rm 
matter}-|\nabla \varphi|^2
-\frac{1}{4} \log \epsilon \frac{M_4^4}{M_5^6} 
\Biggl(  -4R_{\mu\nu} \nabla^\mu \varphi^* \nabla^\nu \varphi 
+\frac{4}{3}R |\nabla \varphi|^2 \nonumber \\
& &~~+ R^\mu_\nu R_\mu^\nu -\frac{1}{3}R^2+\frac{2}{3}|\nabla \varphi|^4
+2|(\nabla \varphi)^2|^2 \Biggr) +2|\nabla^2 \varphi |^2 \Biggr]+ \Gamma_{\rm CFT}, 
\label{action}
\end{eqnarray}
%
where $\Gamma_{\rm CFT}$ is the effective action for the holographic CFT field on the 
brane. $g_{\mu\nu}$, $R$ and $\nabla_\mu$ are the induced metric, the Ricci scalar and 
the covariant derivative on the brane. 
$M_4$ is the four dimensional Planck scales given by $M_4^2 = l M_5^3$ where 
$l$ is the curvature radius of the adS spacetime defined by 
$\Lambda=-6 M_5^3/l^2$. ${\cal L}_{\rm matter}$ is the leading part of 
the Lagrangian density for the other localised matters on the brane. 
The higher derivative terms come from the counter terms \cite{GKR,SKOT}. 
$\epsilon$ determines the renormalization scale of the CFT.

A point is that, even if the scalar field has no potential in higher dimensional 
theory, $R (\partial \varphi)^2$ interaction terms can produce the B-violation 
process in the effective theory. Indeed, for the current 
$J^\mu := -i(\varphi \nabla^\mu \varphi^*-\varphi^* \nabla^\mu \varphi )$, 
its divergence becomes 
%
\begin{eqnarray}
\nabla_\mu J^\mu \simeq \frac{M^4_4}{M_5^6}\Biggl[ \frac{2}{3}(\nabla_\mu R) J^\mu +4 
R_{\mu\nu}
\nabla^\mu J^\nu  \Biggr], \label{current}
\end{eqnarray}
%
where we assumed that the contribution of the scalar fields into the background geometry 
is negligible and neglected a numerical coefficient, $\frac{1}{4}{\rm log}\epsilon$. 
In the expanding universe, $R_{\mu\nu}$ has the time dependence in general. This means 
that $J_\mu$ is not conserved in general. 
It is reminded that there is a 
conserved current $\tilde J^\mu$ by virtue 
of the 
global $U(1)$ symmetry
%
\begin{eqnarray}
\tilde J^\mu & = & J^\mu
- \frac{M_4^4}{M_5^6}\Biggl[4R^\mu_\nu J^\nu
-\frac{4}{3}RJ^\mu-\frac{4}{3} |\nabla \varphi|^2 J^\mu \nonumber\\
&& +4i\Biggl( \varphi^* \nabla^\mu \varphi (\nabla 
\varphi^*)^2 -\varphi \nabla^\mu \varphi^* (\nabla \varphi)^2 \Biggr) \Biggr]. 
\end{eqnarray}
%

\section{Gravitational baryogenesis}
For the generation of the net baryon number, C and CP violations are also required 
\cite{3con}. Let us consider a new C and CP violation source recently proposed in 
paper \cite{GB}(Gravitational baryogenesis). 
The reason why we consider is that the essence is surprisingly common with the new 
B-violation process found in the previous section, that is, 
CP violation is supposed to be raised from the interaction 
%
\begin{eqnarray}
S_{\rm int} \sim \frac{1}{M_*^2} \int d^4x {\sqrt {-g}} (\partial_\mu R) J^\mu.
 \label{int}
\end{eqnarray}
%
Note that there appears the same factor $\partial_\mu R$ as the B-violating
terms. The authors in Ref. \cite{GB} expect such an interaction in the low energy 
effective theory due to some non-perturbative effects from quantum gravity. 
This is the simplest coupling to the spacetime curvature that breaks C and CP.
In the brane world context, such a term may exist on the brane in the same way 
as the B-violating terms. For example, we assume that there is 
an explicit CP violating term on the brane as supposed in 
spontaneous baryogenesis \cite{CK},
%
\begin{equation}
S_{\rm int}= \frac{1}{M} \int d^4 x \sqrt{-g} (\partial_\mu \theta) J^\mu 
\end{equation}
%
where $\theta$ is a real scalar field and $M$ is a constant 
with dimension of mass. 
Then the integration by parts and using of Eq. (\ref{current}) 
bring us the following interaction term
%
\begin{eqnarray}
S_{\rm int} & \simeq & \frac{M_4^4}{M_5^6}
\int d^4 x{\sqrt {-g}} \left( \frac{-\theta}{M} \right) 
\Biggl[\frac{2}{3} (\nabla_\mu R) J^\mu 
+4 R_{\mu\nu} \nabla^\mu J^\nu  \Biggr]. 
\end{eqnarray}
%
This is the curvature-current interaction that leads to 
gravitational baryogenesis.  

Then $M_{*}$ is expected to be proportional to $M^3_5/M^2_4$. 
The coefficient depends on the model for the violation of CP, so we set 
$M_*=f M^3_5/M^2_4$ and leave $f$ as a parameter in our model. 
If $J^\mu$ is a current which produces net $B-L$, 
it is not wiped away by the electroweak sphaleron process \cite{KRS}. 
The interaction (\ref{int}) violates CPT spontaneously 
due to the dynamics of the expanding universe. Thus the out of 
thermal equilibrium is not necessary for the baryogenesis. 
We denote the temperature at which the B-violating process is 
decoupled from thermal equilibrium as $T_D$. 
Then the produced baryon to entropy ratio will be  
%
\begin{eqnarray}
\frac{n_B}{s} \sim \left. \frac{\dot R}{M_*^2 T} \right \vert_{T_D}.
\end{eqnarray}
%
where $n_B$ and $s$ are the net baryon number density 
and the entropy density, respectively.
An essential ingredient to realize this mechanism is 
a non-zero time dependence of $R$.
In the conventional four dimensional general relativity, 
$\dot R=0$ in radiation dominated universe because $T^\mu_\mu=0$, 
where $T_{\mu\nu}$ is the energy-momentum tensor of matters. 
Hence the authors in Ref. \cite{GB} considered the quantum effect, that is, 
trace anomaly introduced the non-zero $T^\mu_\mu$. 
On the other hand, in the brane world considered here, $\dot R \neq 0$ 
will be realised even in the radiation dominated universe 
due to the higher order curvature correction terms 
in the effective theory. 

In the Randall-Sundrum braneworld models \cite{RS,BWreview}, 
using the geometrical projection method \cite{SMS}, 
we can derive the following gravitational equation on the brane
%
\begin{eqnarray}
R_{\mu\nu}-\frac{1}{2}g_{\mu\nu}R=\frac{1}{M_4^{2}}T_{\mu\nu}+\frac{1}{M_5^6}\pi_{\mu\nu}-
E_{\mu\nu}. 
\label{effective}
\end{eqnarray}
%
where $T_{\mu\nu}$ is the energy-momentum tensor on the brane, $\pi_{\mu\nu}:= 
-\frac{1}{4}T_{\mu\alpha}T^\alpha_\nu +\frac{1}{12}TT_{\mu\nu}
+\frac{1}{8}g_{\mu\nu}T^\alpha_\beta T^\beta_\alpha -\frac{1}{24}g_{\mu\nu}T^2$ and 
$E_{\mu\nu}$ is projected Weyl tensor.  Here we assumed that the contribution to the 
gravity is 
dominated by the ordinal matter ${\cal L}_{\rm matter}$. 
Hence the equation of Eq. (\ref{effective}) 
is equivalent with that derived from the effective action of Eq. (\ref{action}). 
For example, $E_{\mu\nu}$ contains the holographic CFT 
stress tensor \cite{SKOT,SI}. Furthermore, $\pi^\mu_\mu$  
corresponds to the trace anomaly of holographic CFT fields which is 
evaluated by $ \frac{1}{{\sqrt {-g}}}\frac{\delta \Gamma_{\rm CFT}}{\delta g_{\mu\nu}} 
g_{\mu\nu}$. 
$T_{\mu\nu}$ obeys the local conservation low $\nabla^\mu T_{\mu\nu}=0$. 

From the trace term of Eq. (\ref{effective}) we obtain  
%
\begin{eqnarray}
R=-\frac{1}{M_4^2}T^\mu_\mu- \frac{1}{M_5^{6}} \pi^\mu_\mu
\end{eqnarray}
%
The second term in right-hand side is higher order corrections to conventional cosmology. 
Now we are thinking of the homogeneous and isotropic universe with perfect fluid. In this 
case 
$R$ and $\dot R$ become
%
\begin{eqnarray}
R=(1-3w) \frac{\rho}{M_4^2}-\frac{1}{6}(1+3w) \frac{\rho^2}{M_5^6}
\end{eqnarray}
%
and
%
\begin{eqnarray}
\dot R = -3(1+w) H \Biggl[\frac{1-3w}{M_4^2}-\frac{(1+3w)\rho}{3M_5^6} \Biggr]\rho,  
\end{eqnarray}
%
where $w=P/\rho$, $H=\dot a /a$ and $a$ is the scale factor of the universe. 

In the very early universe, the radiation is dominated and then $w=\frac{1}{3}$. 
So the contribution from the conventional cosmology does vanish. But, the 
corrections remains and then 
%
\begin{eqnarray}
\dot R = \frac{8}{3}\frac{H \rho^2}{M_5^6} \sim \frac{T^{10}}{M_5^6 M_4},
\end{eqnarray}
%
where $T$ is the temperature. 
Therefore the reaction rate is given by 
%
\begin{eqnarray}
\Gamma_B \sim \frac{M_4^4}{M_5^6} \dot R \sim \frac{M_4^3T^{10}}{M_5^{12}}.
\label{reaction}
\end{eqnarray}
%
The decoupling temperature $T_D$ is the temperature at which 
$\Gamma_B =H$, thus it is determined as  
%
\begin{eqnarray}
T_D \sim \frac{M_5^{3/2}}{M_4^{1/2}} 
\end{eqnarray}
%
In Eq. (\ref{reaction}) we tacitly assumed that the first term of right-hand side in 
Eq. (\ref{current}) is dominated. It is easy to check that $T_D$ does not depend on 
such an assumption. Then the net baryon to entropy ratio is given by
%
\begin{eqnarray}
\frac{n_B}{s} & \sim &  \frac{T_D^9}{M_*^2 M_5^6 M_4} 
\sim  \frac{M_4^3 T_D^9}{f^2 M_5^{12}} 
\sim \frac{1}{f^2} \Biggl(   
\frac{M_5}{M_4}\Biggr)^{3/2}.  
\end{eqnarray}
%
This is quite an interesting result. The resultant baryon to entropy ratio
is determined by the ratio of the Planck scales in four dimensional 
and five dimensional spacetime except for the parameter for CP 
violation.

Now we consider the parameters in our model. In RS model, there is 
one parameter $l$ that determines the scale below which 
Newton's force low is modified. 
Today's experiment restricts $l$ smaller than $0.2$mm. Then $M_5$ is 
bounded below, $M_5 >10^{8}{\rm GeV}$. The tension of the brane is 
also bounded below as $\sigma > ({\rm TeV})^4$. 
The decoupling temperature is estimated as 
%
\begin{eqnarray}
T_D \sim 10^{2.5}
\Biggl( \frac{M_5}{10^8{\rm GeV}} \Biggr)^{3/2}{\rm GeV}
\end{eqnarray}
%
Finally we obtain the net baryon to entropy ratio 
%
\begin{eqnarray}
\frac{n_B}{s} \sim 10^{-10} \Biggl( \frac{0.001}{f} \Biggr)^2 
\Biggl( \frac{ 10^{8}{\rm GeV}}{M_5}\Biggr)^{12} 
\Biggl( \frac{T_D}{ 10^{2.5}{\rm GeV}}\Biggr)^9.  
\end{eqnarray}
%
This is quite reasonable value at $T_D \sim 10^{2.5}{\rm GeV}$, $M_5 \sim 10^8 {\rm GeV}$ 
and 
$f\sim 0.001$. As Ref. \cite{GB}, we should require $T_D < T_R < M_I$ where $T_R$ and 
$M_I$ are 
the reheating temperature and the inflationary scale if the entropy production was 
occurred at the reheating after the inflation. For the chaotic inflation with the 
potential 
$V=\frac{1}{2}m^2 \phi^2$ at the very high energy, observations lead us the constraints 
$m \sim 5 \times 10^{-5} M_5$ and $\phi \sim 3\times 10^2 M_5$\cite{chaotic}. Thus, the 
inflationary 
scale becomes $M_I \sim V^{1/4} \sim 10^{-0.5}M_5$ which satisfies $T_D \ll M_I$.

\acknowledgments{
TS would like to thank Norisuke Sakai and Katsushi Ito for discussions. 
The work of TS was supported by Grant-in-Aid for Scientific
Research from Ministry of Education, Science, Sports and Culture of 
Japan(No.13135208, No.14740155 and No.14102004). The work of KK was 
supported by JSPS.
}
\appendix
\section{The derivation of effective action on the brane}
adS/CFT correspondence in the braneworld can be formulated using the partition function 
in the path integral presentation \cite{GKR} 
%
\begin{eqnarray}
Z  =  \int {\cal D}G {\cal D} \Phi e^{iS_5(G,\Phi)+\frac{i}{2}S_{\rm brane}(g)} 
  =  \int {\cal D}g {\cal D} \varphi e^{\frac{i}{2}S_{\rm brane}+iS_{\rm ct} +i\Gamma_{\rm 
CFT}},
\end{eqnarray}
%
where $\Gamma_{\rm CFT}$ is the effective action for holographic CFT on the brane. 
$\varphi$ is the value at the location of the brane, $\varphi = \Phi|_{\rm brane}$. 
$S_{\rm ct}$ is 
the counter-term which makes the total action finite. $S_{\rm ct}$ can be derived as the 
solution 
to the Hamilton-Jacobi(HJ) equation:
%
\begin{eqnarray}
& & -\frac{2}{{\sqrt {-g}}} \frac{\delta S}{\delta g_{\mu\nu}}  \frac{\delta S}{\delta 
g_{\alpha\beta}}
\Biggl( g_{\mu\alpha}g_{\nu\beta} -\frac{1}{3}g_{\mu\nu} g_{\alpha\beta} \Biggr) 
-\frac{1}{2{\sqrt {-g}}} \Biggl[ 
\Bigl(\frac{\delta S}{\delta \phi} \Bigr)^2
+\Bigl(\frac{\delta S}{\delta \chi} \Bigr)^2 \Biggr] \nonumber \\
& & +\frac{1}{2}{\sqrt {-g}} \Biggl[(\nabla \phi)^2+ (\nabla \chi)^2 
-(R-2\Lambda) \Biggr] =0.
\end{eqnarray}
%
where $\varphi = (1/{\sqrt {2}})(\phi + i \chi)$. 
HJ equation can be solved by gradient expansion with small parameter 
$(l/L)^2$, where $l$ is the adS radius and $L$ is typical length 
scale on the brane \cite{Koyama}.  After all, 
the total effective action on the brane becomes
%
\begin{eqnarray}
S_{\rm eff}=S_{\rm ct}+\frac{1}{2}S_{\rm brane}+\Gamma_{\rm CFT}.
\end{eqnarray}
%
See Refs. \cite{GKR,SKOT,Koyama} for the details.

\end{document}